\documentclass[showpacs,preprintnumbers,amsmath,prl,
graphicx,amssymb,superscriptaddress,twocolumn]{revtex4}
\usepackage{graphicx}
\usepackage{dcolumn}
\usepackage{bm}

\renewcommand{\today}{\number\day\space\ifcase\month\or January\or 
 February\or March\or April\or May\or June\or July\or August\or 
 September\or October\or November\or December\fi\space\number\year}

\begin{document}
\title{Results from a Search for Light-Mass Dark Matter 
with a \\ 
P-type Point 
Contact Germanium Detector}

\newcommand{\pnnl}{Pacific Northwest National Laboratory, Richland,
WA 99352, USA}
\newcommand{\uc}{Kavli Institute for Cosmological Physics and Enrico
Fermi Institute, University of Chicago, Chicago, IL 60637, USA}
\newcommand{\llnl}{Lawrence Livermore National Laboratory,
Livermore, CA 94550, USA}
\newcommand{\canberra}{CANBERRA Industries, Meriden, CT 06450, USA}
\newcommand{\anl}{Argonne National Laboratory, Argonne, IL 60439, USA}
\newcommand{\snl}{Sandia National Laboratories, Livermore, CA 94550,
USA}
\newcommand{\uw}{Center for Experimental Nuclear Physics and
Astrophysics and Department of Physics, University of Washington,
Seattle, WA 98195, USA}
\newcommand{\ornl}{Oak Ridge National Laboratory, Oak Ridge, TN 37831, USA}
\newcommand{\unc}{Department of Physics and Astronomy, University of 
North Carolina, NC 27599, USA}


\affiliation{\pnnl}
\affiliation{\uc}
\affiliation{\llnl}
\affiliation{\snl}
\affiliation{\canberra}
\affiliation{\anl}
\affiliation{\uw}
\affiliation{\ornl}
\affiliation{\unc}
														
\author{C.E.~Aalseth}\affiliation{\pnnl}
\author{P.S.~Barbeau}\affiliation{\uc}
\author{N.S.~Bowden}\affiliation{\llnl}
\author{B.~Cabrera-Palmer}\affiliation{\snl}
\author{J.~Colaresi}\affiliation{\canberra}
\author{J.I.~Collar$^{*}$}\affiliation{\uc}
\author{S.~Dazeley}\affiliation{\llnl}
\author{P.~de Lurgio}\affiliation{\anl}
\author{G.~Drake}\affiliation{\anl}
\author{J.E.~Fast}\affiliation{\pnnl}
\author{N.~Fields}\affiliation{\uc}
\author{C.H.~Greenberg}\affiliation{\uc}
\author{T.W.~Hossbach}\affiliation{\pnnl}\affiliation{\uc}
\author{M.E.~Keillor}\affiliation{\pnnl}
\author{J.D.~Kephart}\affiliation{\pnnl}
\author{M.G.~Marino}\affiliation{\uw}
\author{H.S.~Miley}\affiliation{\pnnl}
\author{M.L.~Miller}\affiliation{\uw}
\author{J.L.~Orrell}\affiliation{\pnnl}
\author{D.C. Radford}\affiliation{\ornl} 
\author{D.~Reyna}\affiliation{\snl}
\author{R.G.H.~Robertson}\affiliation{\uw}
\author{R.L.~Talaga}\affiliation{\anl}
\author{O.~Tench}\affiliation{\canberra}
\author{T.D.~Van Wechel}\affiliation{\uw}
\author{J.F.~Wilkerson}\affiliation{\uw}\affiliation{\unc}
\author{K.M.~Yocum}\affiliation{\canberra}

\collaboration{CoGeNT Collaboration}
\noaffiliation


\begin{abstract}
We report on several features present in the energy spectrum 
from an ultra low-noise germanium detector 
operated at 2,100 m.w.e. By implementing a new technique able to 
reject surface events, a number of cosmogenic peaks can be observed for 
the first time. We discuss several possible causes for an irreducible
excess of bulk-like events below 3 keVee, including a dark matter 
candidate common to the DAMA/LIBRA annual 
modulation effect, the hint of a signal in CDMS,
and phenomenological predictions. Improved constraints are 
placed on a cosmological origin for the DAMA/LIBRA effect.
\end{abstract}

\pacs{85.30.-z, 95.35.+d, 95.55.Vj, 14.80.Mz\\
$^{*}$~Corresponding author. E-mail: collar@uchicago.edu}

\maketitle

We have recently presented first dark matter limits \cite{ourprl} from the underground 
operation of p-type point contact (PPC) germanium detectors. PPCs
display an unprecedented combination of target mass and reduced electronic 
noise, resulting in an enhanced sensitivity to
low-energy rare events. Promising 
applications in astroparticle and neutrino physics are expected from 
this technology \cite{jcap}. 

Sources of detector radiocontamination became evident 
during operation of a PPC at a depth of 330 m.w.e. \cite{ourprl}. 
A new 440 g PPC diode was built with 
attention paid to these. The new crystal is a modification of the 
BEGe (Broad Energy Germanium) geometry, a commercial quasi-planar PPC 
design from Canberra Industries. It is 
operated in the Soudan Underground Laboratory (Minnesota, USA). This 
improved detector, while still not featuring all possible 
 measures against backgrounds (electroformed cryostat components, etc.),
has delivered more than one order-of-magnitude background reduction compared to 
\cite{ourprl}. The background 
achieved below $\sim$3 keVee (keV 
electron equivalent, i.e., ionization energy), down to the 0.4 keVee
electronic noise threshold, 
is so far the lowest reported by any dark matter 
detector. 
The shielding, electronics and basic pulse-shape 
discrimination (PSD) are similar to those described in \cite{ourprl}, 
with the addition of data  
storage of raw preamplifier traces, and removal of two internal 
active shields. An external muon veto is preserved. Veto-coincident 
events do not accumulate
in excess near threshold: we therefore do not apply veto 
cuts to present data, to avoid a dead time 
penalty.

A number of peaks are observed in low-energy 
spectra from ultra-low background germanium detectors. 
They originate in activation 
of the crystal by exposure to cosmogenic neutrons and 
protons
at sea level. Long-lived radioactive 
products result from their spallation of  
germanium nuclei. Whenever this progeny decays 
via electron capture (EC), the deposited energy 
can be limited to the 
atomic binding energy of the  daughter's captured electron, released as 
short-ranged X-rays and Auger electrons. Taking place within 
the crystal, these are detected with $\sim\!100$\% efficiency, giving 
rise to 
the observed peaks. 

\begin{figure}
\includegraphics[width=7.7cm]{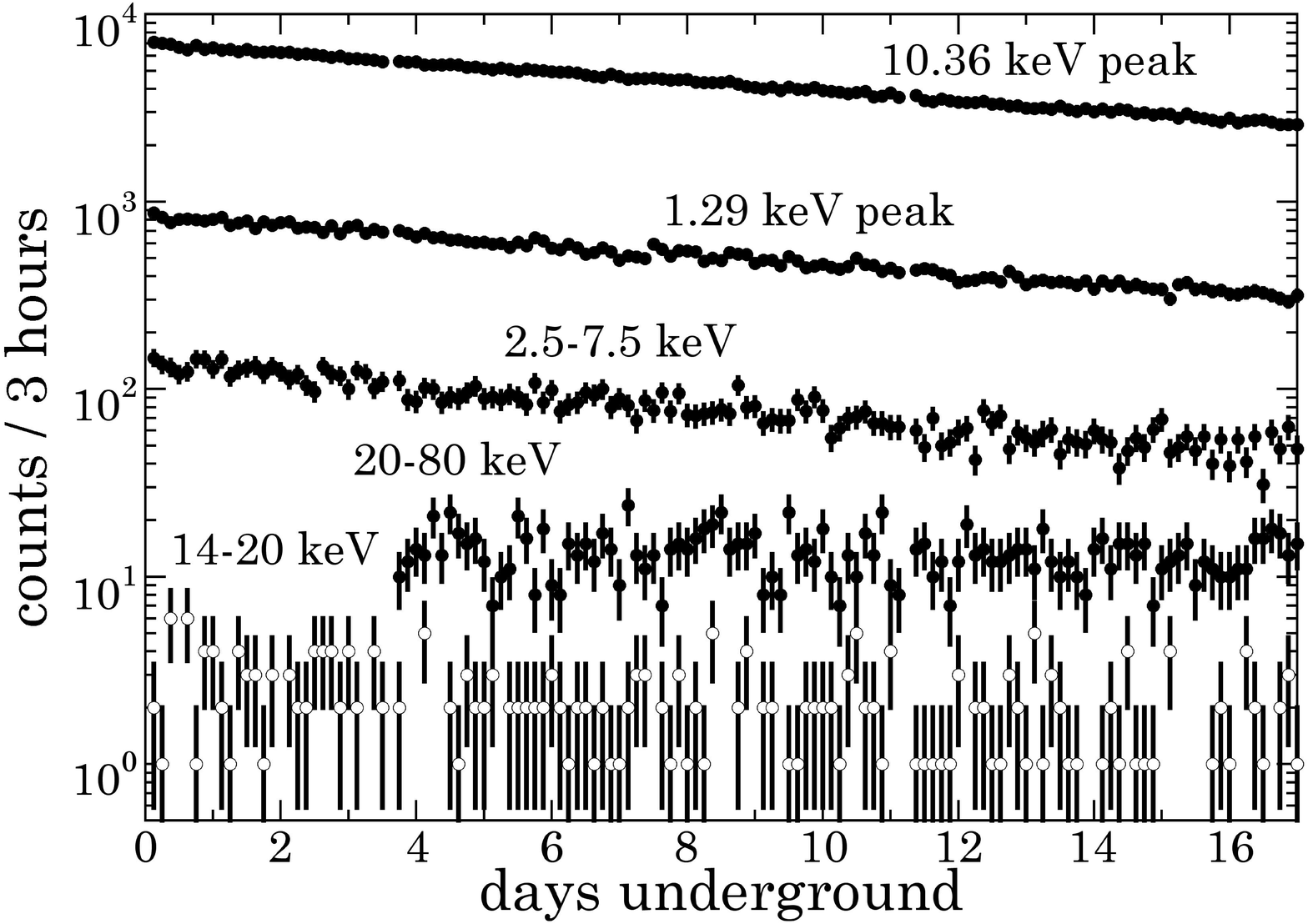}
\caption{\label{fig:epsart}Decays associated with $^{71}$Ge
($T_{1/2}=11.4$d) produced via thermal 
neutron activation of a PPC detector (see text).
}
\end{figure}

Due to the very short attenuation lengths (few microns) expected 
from low-energy X-rays in solids and the exclusive use of radioclean 
materials near the crystals, all low-energy peaks observed so 
far have a cosmogenic origin internal to germanium. For p-type diodes an 
additional
obstacle against external low-energy radiation arises from 
a quasi-inactive n+ contact, spanning most of the surface of the 
semiconductor. This contact is created by lithium diffusion   
down to a depth 0.5-1 mm. 
Fig.\ 1 displays the decay of the 10.36 keV K-shell 
EC peak from $^{71}$Ge produced via intense thermal neutron activation of
a PPC. A peak at 1.29 keV, 
originating in L-shell EC, exhibits the same 
decay (also the region 0.5-1.29 keV, not shown 
for clarity). 
So does the 2.5-7.5 keV ``plateau'', 
but not events above 10.36 keV.  
The ratio of
activity in the plateau
to that under the 10.36 keV peak matches
the estimated fraction of Li-diffused volume, 
suggesting an origin for most plateau events in
partial charge collection from  
$^{71}$Ge decays within the n+ contact. 

\begin{figure}
\includegraphics[width=8.cm]{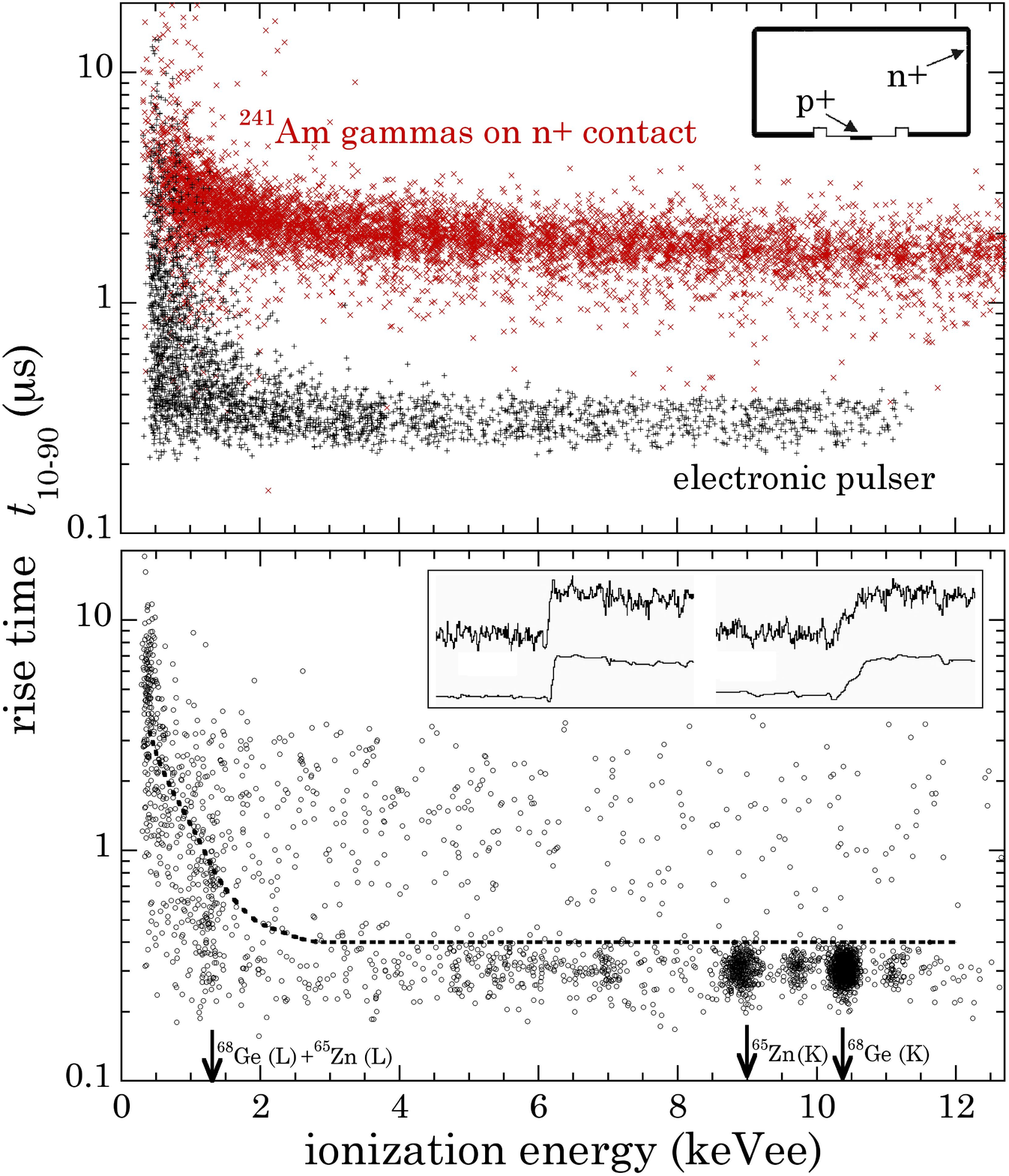}
\caption{\label{fig:epsart}{\it Top panel:} Rise time in preamplifier 
traces from $^{241}$Am gammas and a manual scan of
reference electronic pulser signals (see text). A change 
in digitizer range is noticeable above $\sim4$ keV.
{\it Bottom panel:} {\it Idem} for background
events collected by the PPC in Soudan. The dotted line 
represents the 90$\%$ signal acceptance contour for bulk events.
{\it Top inset:} Vertical cross section of a BEGe PPC detector, 
showing the two contacts. 
{\it Bottom inset:} Typical preamplifier traces from 1 keVee events, 
before and after wavelet denoising, for $t_{10-90}=0.22~ 
\mu s$ (left) and $t_{10-90}=1.53~ \mu s$ (right).}
\end{figure}

These partial energy depositions could be an issue, in that they 
accumulate signals in the region of most interest for dark matter studies, 
i.e., near threshold. In inspecting preamplifier traces from PPCs we noticed a population 
of low-energy slow pulses, featuring rise times 
($t_{10-90}$) significantly longer than the 
typical $t_{10-90}\!\sim\!\!0.3~ \mu s$. These are mentioned in 
early germanium detector literature  as originating 
precisely in the n+ contact. Their cause is the weak
electric field intensity next to the lithium-diffused region \cite{early}. We 
demonstrated the association between partial charge collection and slow rise time  
by irradiating the``closed end'' (side opposite to p+ 
contact, Fig.\ 2 inset) of the PPC in 
\cite{ourprl} with gammas from a $^{241}$Am source. 
Fig.\ 2 displays the much longer rise times associated with partial 
energy depositions in the n+ contact from the short-ranged 
59.5 keV gamma 
(attenuation length in Ge $\sim$1 mm). Full-energy  
depositions, taking place deeper in the crystal, produce the 
expected $t_{10-90}\!\sim0.3~ \mu s$. Using a MCNP-PoliMi \cite{polimi} 
simulation of the energy-depth profile in this calibration, it is 
possible to faithfully reconstruct the 
$^{241}$Am energy spectrum when the charge 
collection efficiency $\epsilon$ is described by a best-fit logistic (sigmoid) function 
of the form $\epsilon=1/(1+43.5~ e^{-86~ (d-0.14)})$, where $d$ is 
the interaction depth in cm. This implies an outermost ``dead'' layer of 
$\sim$1 mm, followed by a $\sim$1 mm ``transition'' layer prone to partial charge 
collection, in good agreement with \cite{early}. Since we intend 
to reject surface events 
in our dark matter search by performing data cuts based on $t_{10-90}$, in what 
follows we 
conservatively revise 
the fiducial mass of our detector to be 330 g (two outer mm 
discarded). 

Based on this discussion, low energy (few keV) radiation can reach the 
PPC active volume through a single region,
the intra-contact passivated surface, at the center of which the 5 mm p+ point-contact is 
established (Fig.\ 2, inset). The protective SiO$_{x}$ passivation 
layer is just $\sim$1500 Angstroms thick. Any events 
arising from an external low-energy
source must originate from materials in the line-of-sight 
of this surface (diameter 2.2 cm). These are specially-etched virgin PTFE,
similarly treated 
OFHC copper and a needle contact (gold-plated 
brass, its tip wetted with low-background pure tin). 

Fig.\ 2 (bottom) shows the 
rise time distribution for low-energy events in the 
PPC at Soudan. These data correspond to an eight week period starting three 
months after underground installation, to allow for nearly 
complete $^{71}$Ge decay. 
The top panel shows the same 
distribution for a collection 
of electronic pulser events in this detector. After a small upwards 
shift in $t_{10-90}$ by 0.1$\mu$s, these strongly resemble 
radiation-induced bulk 
events in this representation. The shift accounts for the additional charge 
collection time affecting energy depositions in the bulk of the 
crystal. Simulations of  
charge collection corroborated the magnitude 
of the applied shift. The dotted 
line in the bottom panel represents the 
90$\%$ boundary for signal acceptance of pulser events. We further 
confirmed
that this signal acceptance also applies to bulk events 
by observing the preservation of the L-shell EC
activity from $^{68}$Ge (1.29 keV) and $^{65}$Zn (1.1 keV), before and after this rise time 
cut. While this 
makes us confident that our signal acceptance for bulk 
events is understood at least 
down to 1 keV, the possibility remains of some unrejected surface 
events closer to threshold. 
A comparison with the distribution of $^{241}$Am surface 
events (Fig.\ 2) indicates that any such contamination should be 
modest.

\begin{figure}
\includegraphics[width=7.7 cm]{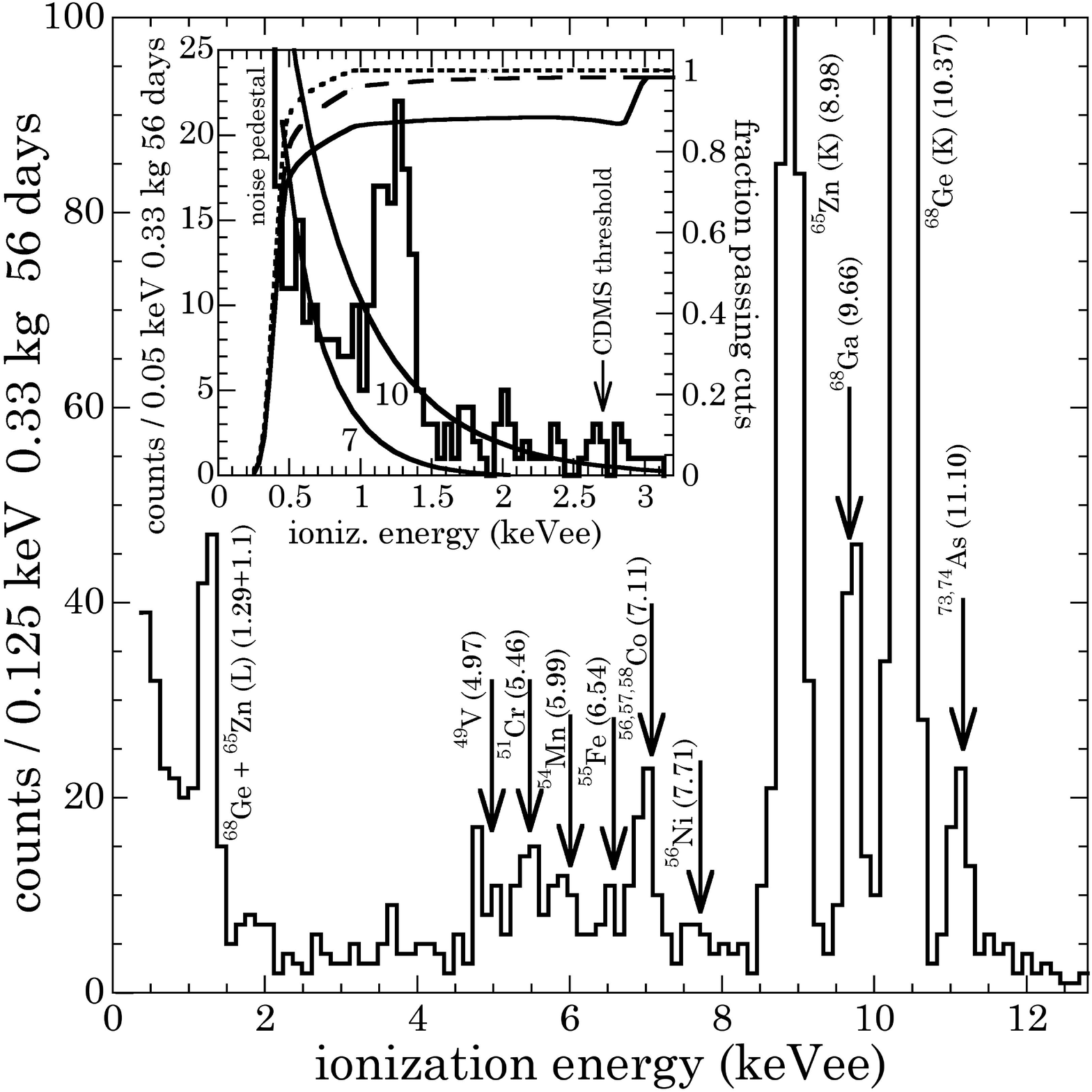}
\caption{\label{fig:epsart} Low-energy spectrum after all cuts, prior 
to efficiency corrections. 
Arrows indicate expected energies for all viable cosmogenic peaks 
(see text). 
{\it Inset:} Expanded threshold region, showing the $^{65}$Zn and 
$^{68}$Ge
L-shell EC peaks. Overlapped on the spectrum are the sigmoids for 
triggering efficiency (dotted), trigger + microphonic PSD cuts (dashed) 
and trigger + PSD + rise time cuts (solid), obtained via 
high-statistics electronic pulser 
calibrations. Also shown are reference signals 
(exponentials) from 7 GeV/c$^{2}$ and 10 GeV/c$^{2}$ 
WIMPs with spin-independent coupling $\sigma_{SI}=10^{-4}$pb.
}
\end{figure}

Fig.\ 3 displays Soudan 
spectra following the rise time cut, which generates a factor 2-3 reduction 
in background (Fig.\ 2). Modest PSD
cuts applied against microphonics are as described in 
\cite{ourprl}. 
This residual spectrum is dominated by events in 
the bulk of the crystal, like those from neutron scattering, 
cosmogenic activation, or dark matter 
particle interactions. Several cosmogenic peaks are noticed, 
many for the first time. 
All cosmogenic products capable of
producing a monochromatic signature are indicated. Observable activities are incipient for all. 

We employ methods identical to those in \cite{ourprl} to 
obtain Weakly Interacting Massive Particle (WIMP)
and Axion-Like Particle (ALP) dark matter limits from these spectra. 
The energy region employed to extract WIMP 
limits is 0.4-3.2 keVee (from threshold to full range of the 
highest-gain digitization channel). 
A correction is applied to 
compensate for signal acceptance loss from cumulative data cuts 
(solid sigmoid in Fig.\ 3, inset). 
In addition to a calculated response function for each WIMP mass \cite{ourprl}, 
we adopt a free exponential plus a
constant as a 
background model to fit the data, with two Gaussians to account for $^{65}$Zn and 
$^{68}$Ge L-shell EC. The energy resolution is as in 
\cite{ourprl}, with parameters $\sigma_{n}$=69.4 eV and $F$=0.29. 
The assumption of an 
irreducible monotonically-decreasing background is 
justified, given the mentioned possibility of a minor contamination 
from residual surface events and the rising concentration 
towards threshold that rejected events exhibit. 
A second source of possibly
unaccounted for low-energy background are the L-shell EC 
activities from observed cosmogenics lighter than $^{65}$Zn. 
These are expected to contribute $<\!15\%$ of the counting 
rate in the 0.5-0.9 keVee region (their L-shell/K-shell 
EC ratio is $\sim\!1/8$ 
\cite{sieg}). A third possibility, quantitatively discussed below, 
consists of recoils from unvetoed 
muon-induced neutrons. 

\begin{figure}
\includegraphics[width=8.cm]{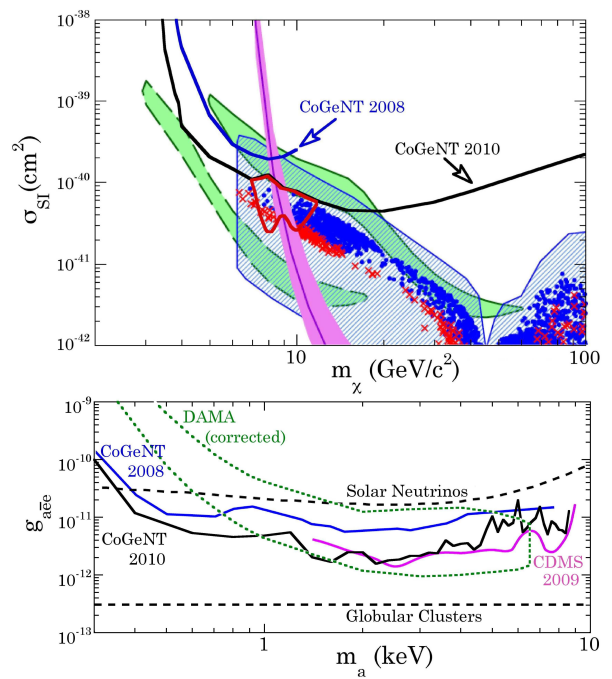}
\caption{\label{fig:epsart}{\it Top panel:} 90\% C.L. WIMP exclusion 
limits from CoGeNT
overlaid on Fig. 1 from \protect\cite{bottino}: green 
shaded patches denote the phase space favoring the
DAMA/LIBRA annual modulation (the dashed contour includes ion 
channeling). Their exact position has been subject to revisions 
\protect\cite{dan2}. The violet band is the region supporting the 
two CDMS candidate events. The scatter plot and the blue hatched 
region represent the supersymmetric models in \protect\cite{bottino2} 
and their uncertainties, respectively. Models including WIMPs with
$m_{\chi}\sim$7-11 GeV/cm$^{2}$ provide a good fit to CoGeNT data
(red contour, see text). The relevance of XENON10 
constraints in this low-mass region has been questioned 
\protect\cite{dan}. {\it Bottom 
panel:} Limits on axio-electric 
coupling $g_{a\bar e e}$ for pseudoscalars of mass $m_{a}$ composing a dark
isothermal galactic halo (see text). 
}
\end{figure}

Fig.\ 4 (top) displays the extracted sensitivity in 
spin-independent coupling ($\sigma_{SI}$) vs.\ WIMP mass 
(m$_{\chi}$). For m$_{\chi}$ in the range $\sim$7-11 GeV/c$^{2}$ 
the WIMP contribution to the model acquires a finite value 
with a 90\% confidence interval incompatible with zero. The 
boundaries of this interval define the red contour in Fig.\ 4.
However, the null hypothesis (no WIMP component in the model) fits the 
data with a similar reduced chi-square $\chi^{2}/dof=$20.4/20 
(for example, the best fit for m$_{\chi}=$ 9 GeV/c$^{2}$ 
provides $\chi^{2}/dof=$20.1/18 
at $\sigma_{SI}=6.7\times10^{-41}$cm$^{2}$). It has been recently
emphasized \cite{bottino} that light WIMP models 
\cite{ourprl,bottino2,bottino3} 
provide a common explanation to the DAMA/LIBRA annual modulation effect 
\cite{damaclaim} and the modest excess of signal-like events  
in CDMS \cite{cdmslast}. 
These WIMP candidates are also compatible with
CoGeNT data (Fig.\ 4). When 
interpreted as WIMP interactions, 
the low energy (3.3 and 4.2 keVee) of the
CDMS recoil-like events is suggestive of a 
light WIMP. However, the 2.7 keVee CDMS threshold
(Fig.\ 3, inset) does not allow for a ready identification of 
such candidates.

Fig.\ 4 (bottom) shows limits on axioelectric dark matter couplings, extracted 
as in \cite{ourprl} from the 
region 0.5-8.5 keVee. Sensitivity to this coupling should improve with additional 
exposure. A discussion on the relevance of the 
DAMA/LIBRA-favored 
region in this phase space is provided in \cite{collarmarino}. 
Lastly, we refer to \cite{ourprl} for a criticism of
ion channeling (Fig.\ 4) as part of the DAMA/LIBRA effect. In light 
of our improved sensitivity, a fair treatment of this 
possibility (i.e., including channeling also for germanium crystals) should render 
it highly constrained.

Enticing as it is to contemplate cosmological implications from 
low-energy spectral features, our focus must remain  
on finding explanations based on natural 
radioactivity. One evident 
possibility is a contribution from recoils caused by environmental or muon-induced 
neutrons. MCNP-PoliMi transport of the environmental neutron flux at 
Soudan \cite{soudanneutron}
through our 
shielding geometry generates a prediction short of the observed rise 
by two orders of 
magnitude. This prediction matches other studies performed at 
the same underground depth \cite{gloria}. The muon-induced 
contribution prior to vetoing is 
simulated following \cite{gloria,mei}, and is found to yield 
just $7\%$ of the rate at threshold (the observed muon veto
coincidences limit this process to 
$<15\%$). Partial energy depositions from high 
energy gammas are not expected to accumulate in this region \cite{compton}. Other 
possibilities, including alpha-recoils from radon deposition on the 
SiO$_{x}$ passivated surface, degraded beta emissions from $^{40}$K in PTFE, 
excess $^{210}$Po \cite{ron} on the 
specially-selected pure tin wetting the central 
contact, etc., have been studied and found lacking. The hypothesis that 
the rise might be due to unrejected electronic 
noise would involve a dramatic deviation from its expected behavior 
near threshold \cite{noise}. Individual inspection of pulses 
comprising the rise reveals them as asymptomatic (Fig.\ 2, inset). 
One conjecture that merits
serious investigation is the effect of a surface channel 
on the intra-contact surface \cite{channel}, possibly leading 
to a population of degraded 
energy events that might nevertheless escape rise time cuts.

In conclusion, we presently lack a satisfactory 
explanation for the observed low-energy rise in a PPC spectrum devoid 
of most surface events. In view of its apparent agreement with 
existing WIMP models,
a claim and a glimmer 
of dark matter detection in two other experiments, it is tempting 
to consider a cosmological origin.
Prudence and past experience prompt us to continue work to
exhaust less exotic possibilities. We extend an invitation to other researchers 
in this field to proceed with the same caution. If this 
feature were to originate in dark matter interactions, a 
PPC-based 60 kg {\sc{Majo\-ra\-na}} 
Demonstrator \cite{majorana} 
would unequivocally detect an annual modulation effect 
in both rate and average energy deposited. Indirect searches should soon probe the 
relevant WIMP phase space region \cite{dan2}.

\section{ACKNOWLEDGMENTS}
This work was sponsored by NSF grants PHY-0653605, PHY-0239812, 
PHY-0114422, LLNL contract DE-AC52-07NA27344, Sandia and Pacific Northwest 
National Laboratories LDRD programs, 
and the Office of Nuclear Physics, U.S. DOE. 
N.F. is supported by the DOE/NNSA SSGF program.
SAND Number: 2010-1375J, LLNL-JRNL-425007. 
We are much indebted to the personnel at the Soudan Underground 
Laboratory and our 
{\sc{Majo\-ra\-na}} and GERDA colleagues.

\end{document}